# MECHANICAL DESIGN OF A HIGH ENERGY BEAM ABSORBER FOR THE ADVANCED SUPERCONDUCTING TEST ACCELERATOR (ASTA) AT FERMILAB*

C. Baffes**, M. Church, J. Leibfritz, S. Oplt, I. Rakhno, FNAL, Batavia, IL 60510, USA

*Abstract*

A high energy beam absorber has been built for the Advanced Superconducting Test Accelerator (ASTA) at Fermilab. In the facility's initial configuration, an electron beam will be accelerated through 3 TTF-type or ILC-type SRF cryomodules to an energy of 750MeV. The electron beam will be directed to one of multiple downstream experimental and diagnostic beam lines and then deposited in one of two beam absorbers. The facility is designed to accommodate up to 6 cryomodules, which would produce a 75kW beam at 1.5GeV; this is the driving design condition for the beam absorbers. The beam absorbers consist of water-cooled graphite, aluminum and copper layers contained in a helium-filled enclosure. This paper describes the mechanical implementation of the beam absorbers, with a focus on thermal design and analysis. The potential for radiation-induced degradation of the graphite is discussed.

## Introduction

Fermi National Accelerator Laboratory is constructing the Advanced Superconducting Test Accelerator (ASTA), a 750MeV electron linac intended to develop and test the technology that will be required for next-generation high-intensity linear accelerators, such as Fermilab's Project X [1] and the ILC. ASTA is housed within the SRF Accelerator Test Facility [2], which, when complete, will also incorporate a cryogenic plant, cryomodule testing facilities, and space for planned future experiments. ASTA will consist on an electron gun, injector, 3 TTF-type or ILC-type cryomodules, downstream diagnostic and experimental beam lines, and a high energy beam dump. The beam dump contains two beam absorbers; one for each of the experimental beam lines. The absorbers are designed to accommodate a future upgrade to the facility, in which a beam would be accelerated through 6 cryomodules. The beam parameters provided by this upgraded machine drive the absorber design, and are shown in the table below:

Table 1: 6-Cryomodule ASTA Key Beam Parameters

| Species | e- |
|---|---|
| Beam Energy | 1.5 GeV |
| Bunch Charge | 3.3 nC |
| Pulse Characteristics | 1 ms duration @ 5Hz |
| Bunches per Pulse | 3000 |
| Total Average Power | 75 kW |
| Beam Transverse Size | Dia 3mm RMS (σ=1.5mm) |

## Absorber Configuration

The beam absorbers must terminate the beam and reject its power to cooling water. The high thermal load, radiation environment, and the need to achieve high reliability were key considerations in the design. A section view of an absorber is shown in Figure 1. The beam exits the beamline through a vacuum window, and is then incident on a stack of graphite with transverse dimensions 0.2 X 0.2m and length 0.9m along the beamline. Downstream of the graphite are aluminum, copper and steel "core" plates. The total length of the absorber along the beam direction is 2m. The graphite and the aluminum core plates are sandwiched between water-cooled gun-drilled aluminum cooling plates. All components of the system are conductively coupled to these cooling plates. The most sensitive components are the aluminum core plates just downstream of the graphite; an indium foil was used in this location to ensure good thermal coupling to the cooling plates. Beam power is rejected through the cooling plates to a radioactive-water closed-loop cooling skid circulating 30 gallons per minute through each absorber.

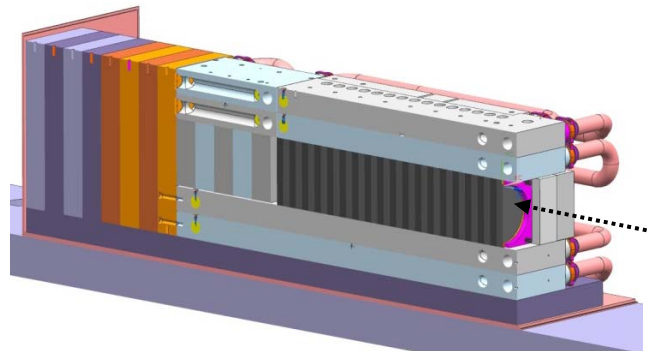

Figure 1: Absorber Configuration: graphite shown in dark grey, aluminum shown in light gray/blue, copper shown in orange, steel shown in purple. Beam enters from right.

In order to prevent oxidization of the graphite and the formation of corrosive compounds due to beam interaction with air, the absorbers are housed within stainless steel enclosures filled with helium at a slight

---

positive pressure relative to atmosphere.

The interaction of beam with the absorbers generates significant prompt and residual radiation. In order to protect the facility and its occupants, the outer layers of the beam dump consist of concrete and steel shielding with overall outside dimensions of 7 X 6 X 7m, and a total weight of approximately 1200 tons.

## Graphite Radiation Damage

During exposure to the beam, graphite will suffer damage due to dislocation of atoms within the material lattice. At damage levels above 0.6 Displacements per Atom (DPA) due to proton beam irradiation, Fermilab has experienced structural failure of graphite [3]. However, significant changes in material properties can occur at much lower damage levels. Most significant in this application are changes in physical size and thermal conductivity.

When used in radiation environments, isotropic graphite has been shown to shrink up to a threshold damage level, and then swell after continuing damage. For the type of isotropic graphite employed in the absorber, data generated from fission applications (with damage dominated by neutrons) indicate that this transition from shrinking to swelling occurs at a damage level of 15 DPA after an accumulated volumetric shrinkage of approximately 7% [4]. In the absorber application, damage levels will be much lower, and the material would be expected to remain in the shrinking regime. In order to maintain thermal contact if any shrinking occurs, the aluminum cooling plates are spring loaded onto the graphite using a stack of Belleville washers that will limit preload variation due to either graphite radiation shrinkage or thermal expansion.

Radiation damage is also known to affect the thermal conductivity of graphite. After irradiation, thermal conductivity can drop to a small fraction of its nominal value. This degradation is the result of vacancies in the lattice interrupting phonon transport, particularly at grain boundaries [5]. At higher temperatures, the material is able to self-anneal to some extent, and thermal conductivity can be recovered. As such, the magnitude of thermal conductivity reduction has a complex dependence on the irradiation and temperature history.

Given the severity of this effect, graphite damage was modeled using the MARS code [6]. Damage was calculated in units of DPA per incident electron. Initial analyses revealed the need to migrate the beam over the absorber in order to avoid local areas with high damage. Integrating over the planned migration profile and a very conservative design particle fluence of 1.4E23 electrons (20 years of full-intensity operation with 70% uptime), a 3-Dimensional damage map for the graphite core was developed (see Figure 2).

In order to perform thermal analysis of a degraded graphite core, it was necessary to define material properties for damaged graphite to use in the analysis. Much data exists in the literature, (e.g. [7]), albeit for damage conditions dominated by fission neutrons. The data may not be representative in this case. In order to conservatively bound existing data, an envelope approach was used. First, literature data were plotted as a thermal conductivity reduction factor: a ratio of damaged thermal conductivity (at a given temperature) to nominal thermal conductivity at the same temperature. Then, five material damage bins were defined. For each bin, a thermal-conductivity-reduction-factor vs. temperature curve was drawn below (i.e. at lower thermal conductivity) the data points within the given damage bin. This is shown in Figure 3. In the ANSYS thermal model, a unique material was defined for each damage bin, with this thermal conductivity reduction factor applied.

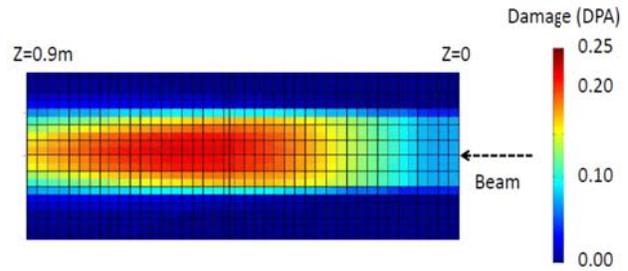

Figure 2: Damage accumulation in graphite core after 20 years of operation at 70% up-time. 0.22 DPA max

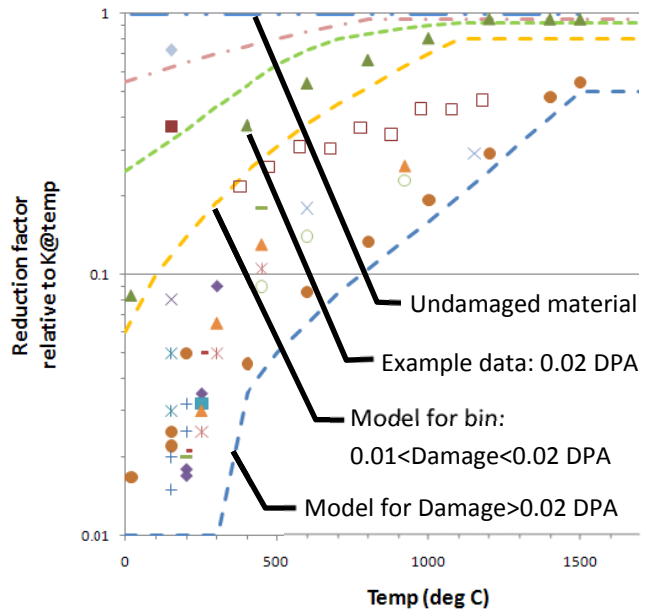

Figure 3: Thermal conductivity reduction factor as a function of temperature and damage. Lines represent factors used in analysis material definitions. Points represent literature data

An interpolation routine was developed to map MARS damage data onto the ANSYS finite element model. Using the 3-dimensional MARS damage estimate as an input, the interpolation routine assigned one of the five discrete material definitions to each element in the graphite ANSYS mesh, on an element-by-element basis. This permitted a more realistic representation of the spatial distribution of thermal conductivity degradation. This damaged-graphite finite element model was used for "End Of Life" thermal analyses, which were compared to "Beginning Of Life" results.

## Energy Deposition and Thermal Analysis

MARS was used to calculate the spatial distribution of energy deposition within the absorber. Results for the steady state case (i.e. energy deposition temporally averaged over many pulses) are shown in Figure 4. As with damage estimates, MARS energy deposition results were interpolated and mapped element-by-element onto the thermal model using volumetric-heating body force commands.

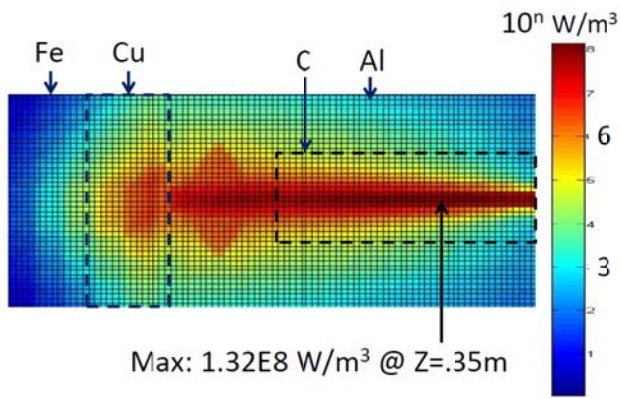

Figure 4: Temporally-averaged energy deposition, logarithmic color scale.

Water cooling was modeled using a uniform convection boundary coefficient on the interior of cooling channels. The convection coefficient of 6400 W/m$^2$K was calculated using the Gnielinski empirical correlation [8]. For reliability and redundancy, each absorber incorporates two redundant cooling circuits. However, the analysis assumes only one circuit in operation.

Several analyses were run, including steady-state, pulse-induced transient, and accident-scenario transient cases. For the purposes of illustrating the effect of graphite damage, steady-state cases are compared at Beginning and End of Life. In "steady-state" cases, temporal-average energy deposition values are used, i.e. the pulsed nature of the beam is neglected. This produces average temperatures, but does not capture peak temperatures near the beam interaction region.

In the Beginning of Life analyses, the graphite assumes nominal, undamaged properties. Maximum temperatures of 640°C are predicted in the graphite. These temperatures, though modest as compared to graphite's sublimation temperature of approximately 2000°C, are high enough to drive the need for inert atmosphere around the absorber. In the End of Life case, graphite temperatures climb to 1700°C. Beginning and End of Life temperature profiles are compared in Figure 5. The extremely low thermal conductivity near the damaged core of the graphite permits a local area to develop high temperatures and high temperature gradients. However, temperatures in areas away from the damaged core of graphite are not greatly affected.

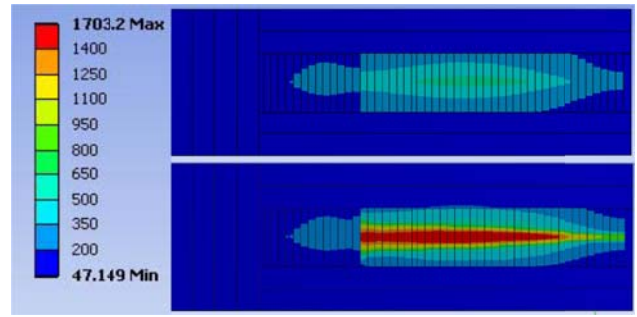

Figure 5: Absorber Temperatures at Beginning of Life (top, max temp. 640°C) and End of Life (bottom, max graphite damage 0.22 DPA, max temp.1700°C)

## Status

The high energy absorbers were installed in late 2011. In the second half of 2012, they will be commissioned in preparation for receiving first beam.

## Acknowledgements

The authors would like to acknowledge M. Cooper, C. Exline, D. Franck, L. Harbacek, W. Johnson, R. Kellett, C. Rogers, and J. Wilson for their efforts in building, testing and installing these devices.

## References

[1] S. Nagaitsev, "Project X – A New Multi-Megawatt Proton Source At Fermilab," PAC '11, New York, USA, 2011
[2] J. Leibfritz et al., "Status and Plans for a Superconducting RF Accelerator Test Facility at Fermilab," IPAC 2012, New Orleans, USA, 2012
[3] Kris Anderson, Fermilab, Private Communication
[4] S. Ishiyama et al., "The effect of high fluence neutron irradiation on the properties of a fine-grained isotropic nuclear graphite," Journal of Nuclear Materials 230 p1, 1996
[5] L.L. Snead, "Accumulation of thermal resistance in neutron irradiated graphite materials," Journal of Nuclear Materials 381 p76, 2008
[6] N.V. Mokhov, "The MARS Code System User's Guide", Fermilab-FN-628, 1995
[7] L.L. Snead and T.D. Burchell, "Thermal conductivity degradation of graphites due to neutron irradiation at low temperature," Journal of Nuclear Materials 224 p222, 1995
[8] V. Gnielinski, "New equations for heat and mass transfer in turbulent pipe and channel flow," Int. Chem. Eng. 16 p359, 1976